\documentclass[aps,prb,twocolumn,showpacs,superscriptaddress,floatfix,10pt]{revtex4-1}

\usepackage{graphicx}
\usepackage{amsmath}
\usepackage{epsfig}
\usepackage{helvet}
\usepackage{amssymb}
\usepackage{url}
\usepackage[pdftex,plainpages=false,bookmarksnumbered=true,hypertexnames=false,colorlinks=true,linkcolor=blue,urlcolor=blue,citecolor=blue,anchorcolor=green]{hyperref}

\def\_{{\textunderscore}} 

\begin{document}

\title{{{TensorTrace:}} an application to contract tensor networks}

\author{Glen Evenbly}
\affiliation{School of Physics, Georgia Institute of Technology, Atlanta, GA 30332, USA}
\email{glen.evenbly@gmail.com}
\date{\today}

\begin{abstract}
Tensor network methods are a conceptually elegant framework for encoding complicated datasets, where high-order tensors are approximated as networks of low-order tensors. In practice, however, the numeric implementation of tensor network algorithms is often a labor-intensive and error-prone task, even for experienced researchers in this area. \emph{TensorTrace} is application designed to alleviate the burden of contracting tensor networks: it provides a graphic drawing interface specifically tailored for the construction of tensor network diagrams, from which the code for their optimal contraction can then be automatically generated (in the users choice of the MATLAB, Python or Julia languages). \emph{TensorTrace} is freely available at \url{https://www.tensortrace.com} with versions for Windows, Mac and Ubuntu.
\end{abstract}

\maketitle

\section{Introduction} \label{sect:Intro}
At a fundamental level, tensor networks serve as compressed representations of correlated data. Although developed in the context of representing quantum and classical many-body systems\cite{TN1,TN2,TN3,TN4}, where they constitute an important class of numeric algorithms, tensor networks have since found application in a diverse range of settings including holography\cite{App1}, machine learning\cite{App2} and big data analysis\cite{App3}. 

The key operation common to all tensor network algorithms is the contraction of a network containing multiple tensors into a single tensor. This is required not only for extracting information from tensor networks, but also for tasks such as optimization of a network (e.g. in order to approximate the ground state of a many-body system). Conceptually, a contraction involves taking the summation over all of the internal indices in a network, but in practice is most efficiently implemented as a sequence of pairwise (or binary) tensor contractions. The numerical implementation of a tensor network algorithm usually involves the following steps for each network required by the algorithm: (i) making a labeled diagram (which serves as design blue-print), (ii) finding the optimal contraction order (or sequence of binary contractions), and (iii) writing the contraction code. Currently, most tensor network practitioners perform each of the tasks above separately, either manually or through the aid of a dedicated program. In particular, there exists many tensor toolboxes designed to help with the third task\cite{TB1,TB2,TB3,TB4,TB5,TB6}. However this paradigm, and specifically the manual translation between blueprint diagrams and computer code, has several significant issues: namely it is both labor-intensive and error-prone. For instance, the optimization of a binary MERA\cite{MERA1,MERA2,MERA3} for the ground state of a $1D$ quantum system requires a total of 14 networks to be programmed (describing the ascending and descending superoperators, as well as the environments of disentanglers and isometries), where each network contains 11 tensors. An error in the coding of any of these networks, such as the improperly ordering of indices on a single tensor, will result in bugs in the algorithm, which can be very difficult to track down. More sophisticated tensor network ansatz, such as PEPS\cite{PEPS}, 2D MERA\cite{2DMERA} or branching MERA\cite{branchMERA}, can require a higher number of even more complicated contractions. This complexity represents a formidable obstacle for new-comers to get started with tensor network algorithms, and also hinders the progress of experienced researchers in the development and application of tensor network methods.

\begin{figure}[!t!b]
\begin{center}
\includegraphics[width=8.5cm]{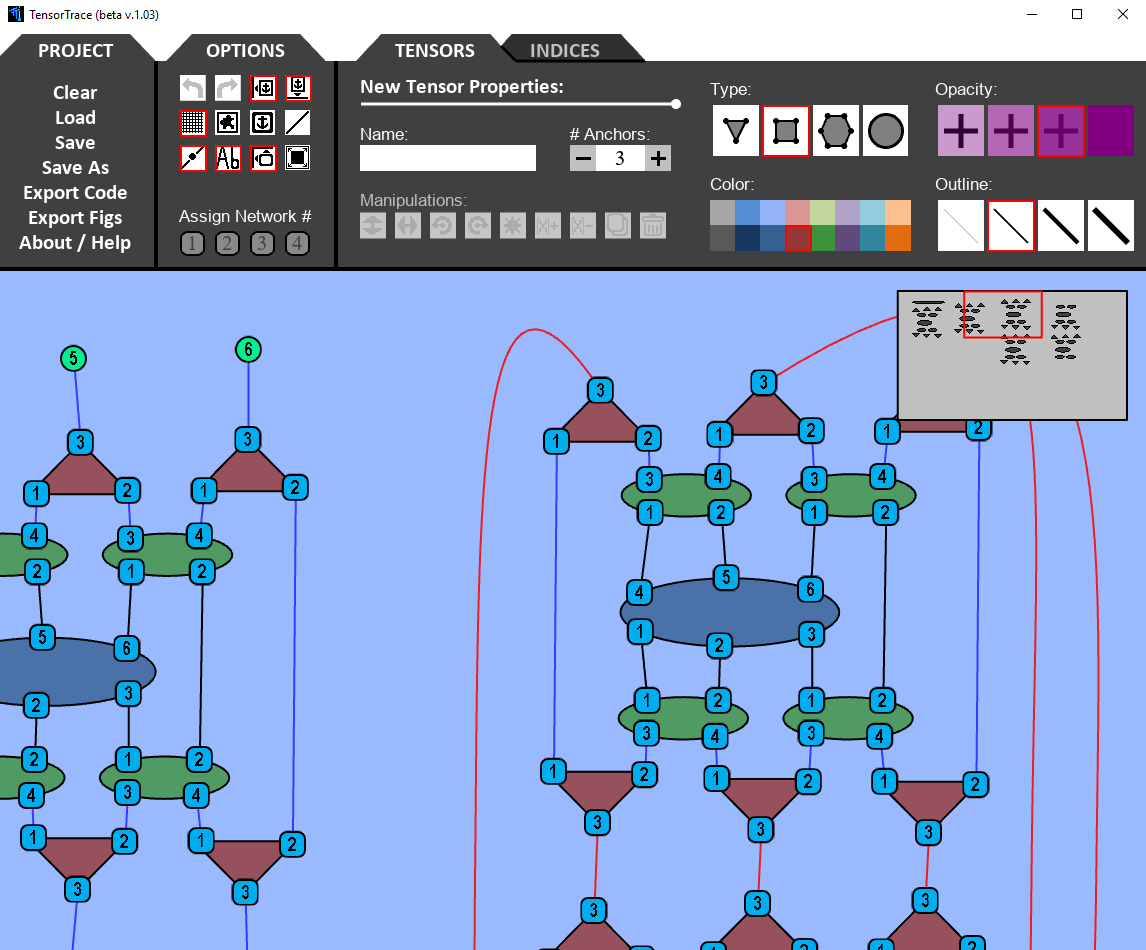}
\caption{Screenshot of the \emph{TensorTrace} interface, showing several tensor network diagrams that have been created.} \label{fig:screenshot1}
\end{center}
\end{figure}

The \emph{TensorTrace} software is intended to simplify the implementation of tensor network algorithms by integrating the three tasks outlined above (i-iii) within a single, purpose-built program. It does this by providing a custom drawing GUI, specifically designed to blue-print tensor network diagrams, from which optimal contraction orders can be automatically determined, and the contraction code automatically generated (in the users choice of MATLAB, Python or Julia languages). Thus, \emph{TensorTrace} can not only reduce the labor involved in producing tensor network code, but also removes a major source of potential errors resulting from translating blue-print diagrams into code.

This manuscript describes how \emph{TensorTrace} can be utilized and provides some examples of its use, as well as details how the code works internally. More information and guides can be found at the dedicated website, \url{https://www.tensortrace.com}, where the program itself can be freely downloaded.

\begin{figure}[!t!b]
\begin{center}
\includegraphics[width=8.5cm]{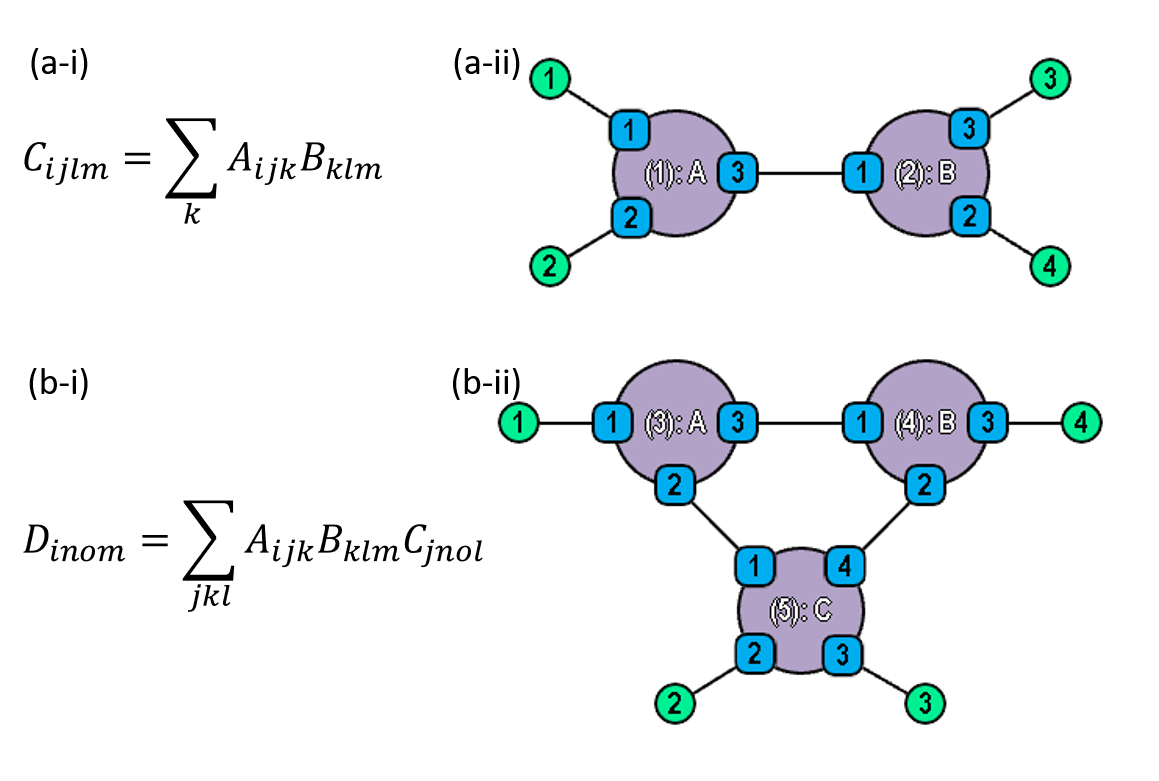}
\caption{(a) Index notation of a contraction between two tensors $A$ and $B$, which makes use dummy labels to indicate index orders, together with the diagrammatic notation used by \emph{TensorTrace} to represent the same contraction. \emph{TensorTrace} uses numbered plaques within each tensor to convey the information about index ordering (e.g. that the third index of $A$ connects with the first index of $B$), whilst the order of indices on the final product tensor is specified via the green plaques appearing on the free ends of open indices. (b) Index notation and the equivalent diagrammatic notation for a contraction involving three tensors.} \label{fig:combnet}
\end{center}
\end{figure}

\section{Diagrammatic notation} \label{sect:Diagram}
\emph{TensorTrace} uses the diagrammatic notation common for this field to represent tensor networks: individual tensors are drawn as solid shapes, while tensor indices are represented as straight or curved lines. For instance a matrix, or $2^\textrm{nd}$-order tensor, is a solid shape with two lines attached. A index connecting between two tensors in the network (called an internal index) denotes a summation over this index, while an index that connects only to a single tensor and has a free end (called an open index) represents one of the indices that appear on the final product tensor (i.e. given by contracting the network). Additionally, \emph{TensorTrace} uses blue numbered plaques within each tensor (which we refer to as an `anchors') which are used to specify the order of indices on that tensor. Similarly the free ends of open indices are also given green numbered plaques, which serve to represent the index ordering on the final product tensor. Thus the diagrams in \emph{TensorTrace} provide complete information about a tensor network, including the order of indices for each tensor, whereas in traditional diagrammatic notation the index ordering conventions are usually specified separately.

Two examples comparing equation notation for tensor networks, where each tensor index is given a dummy label and the network is expressed as a summation over indices, versus the diagrammatic notation are presented in Fig. \ref{fig:combnet}. 

\section{Creating and exporting networks}
\emph{TensorTrace} provides a graphical interface for creating tensor networks, which functions similarly to many common drawing programs. The properties of a tensor (shape, color, opacity, line thickness, number of anchors) can be set in the ``Tensors" tab of the menu, and tensors are created by simply left-clicking and dragging in the main window. The properties of a tensor (or tensors), including the number of anchors present, can also be altered after their creation by first selecting them then changing their properties in the ``Tensors" tab. Tensors may also be given names, which will correspond to the variable names that will be used within generated code (thus should be chosen as valid variable names within the intended programming language). Two or more tensors can be assigned the same name, which will then be treated as copies of the same tensor within the contraction code. If tensor is not given a name then it will be assigned a default name based on the network it belongs to and its order within the network (e.g. the fourth tensor of the second network will be assigned the variable name ``N2\textunderscore 4" by default). 

Index properties are set via the ``Indices" tab of the menu. Each \emph{TensorTrace} project can use up to four types of indices (although this could be increased in future versions of the software). Each index type can be given a name, a default dimension, a set color and a set thickness. Clicking on a tensor anchor in the main window will create an index of the currently selected type, then the index can be completed by clicking on a different anchor (for an internal index) or any free space in the main window (for an open index). Indices are, by default, created in a straight line joining the two anchor points although can be curved by clicking and dragging the index control point. The label that will be assigned to each open index plaque is shown in the top menu, although this label can also be reassigned after the creation of an open index (either by selecting the index and changing in the menu, or by via the keyboard number keys 1-9 when the cursor is hovering over the open index plaque).

\begin{figure}[!t!b]
\begin{center}
\includegraphics[width=8.5cm]{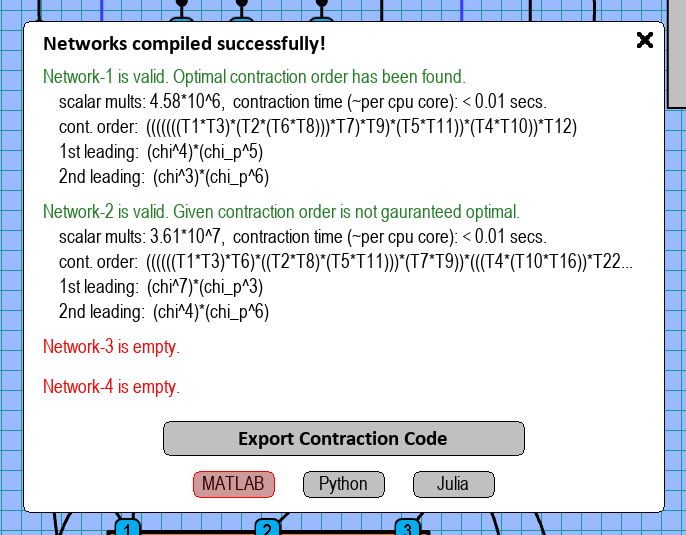}
\caption{Example results produced after choosing the ``Export Code" option on a project containing a pair of tensor networks.} \label{fig:results}
\end{center}
\end{figure}

Once the desired networks are complete, they are then ready to be exported to code. First each of the networks to must be assigned a network number (1-4), which is done by selecting each network then using the appropriate ``Assign Network \#" button in the top menu. A single \emph{TensorTrace} project can contain up to four distinct networks (although the number of networks permissible could also be increased in future versions of the software). Tensors that are unassigned to networks will be ignored by the compiler. The exporting process is then begun by selecting the ``Export Code" option under the ``PROJECT" tab of the menu. Each of the defined networks will then be analyzed by \emph{TensorTrace} to determine whether the network is valid and if so, then to search for the optimal contraction order. The algorithm for determining optimal order is based on that of Ref. \onlinecite{NCON2}, although also incorporates ideas from Ref. \onlinecite{NCON3} to improve the efficiency, and is essentially a brute force search that uses heuristics to rule out some non-optimal solutions as the search progresses. The search algorithm uses the default index dimensions specified in ``Indices" tab of the menu to determine the optimal contraction orders (which may differ depending on the assigned index dimensions). Optimal contraction orders for networks containing $N\le 10$ tensors can typically be determined in a fraction of a second. If any single network contains $N>10$ tensors then \emph{TensorTrace} will present the option to restrict the scope of the search, allowing the user to select \emph{quick}, \emph{thorough} or \emph{extensive} searches. In this restricted setting the algorithm will check only the contraction orders that are most likely to be optimal. Thus the output contraction order may not be guaranteed to be optimal, although is often still likely to be optimal in practice (with the likelihood of optimality increasing as the scope of the search is increased).

After \emph{TensorTrace} has finished analyzing the networks the results will be displayed on screen, as per the example depicted in Fig. \ref{fig:results}. For each network the following information is given: (i) whether the network is valid or invalid, (ii) whether the contraction order given is guaranteed optimal, (iii) total number of scalar multiplications required to contract the network, (iv) an estimation of the time required to contract each network (based on a single 3GHz CPU), (v) the cheapest contraction order found, (vi) the cost of the most expensive and second most expensive binary contractions (expressed as powers of the tensor indices involved). The user is then presented the option of exporting the code (to either of the MATLAB, Python or Julia languages), which will then generate a function file containing the contraction code in the specified language. The contents of these function files and how they are used will be discussed in the next section of this manuscript.

\begin{figure}[!t!b]
\begin{center}
\includegraphics[width=6cm]{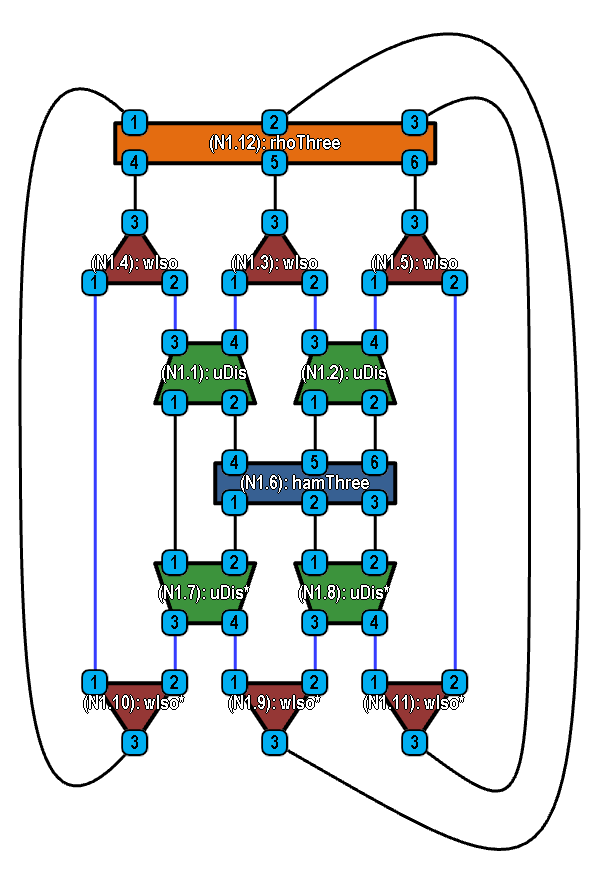}
\caption{One of the networks in the example project ``binaryMERA.ttp" included with \emph{TensorTrace}, which can be exported to produce the code required to optimize a binary MERA for the ground state of a $1D$ quantum system.} \label{fig:binaryMERA3}
\end{center}
\end{figure}

\section{Utilizing the exported code}
After selecting the ``Export Code" option on a project, the user can save a function file in their specified programming language (MATLAB, Python or Julia). The contents of this file is a single function, where the name of the function matches that of the filename chosen by the user. In this section we describe how these functions can be utilized, focusing on the ``binaryMERA" example project included with \emph{TensorTrace} for simplicity, see Fig. \ref{fig:binaryMERA3}, which contains the diagrams necessary to implement variational energy minimization on a binary MERA. Note that the functions generated from \emph{TensorTrace} require the ``ncon" general purpose network contractor\cite{NCON1} to be in working directory, which is included in the \emph{TensorTrace} download and also available at the \emph{tensors.net} website\cite{TenNet}. The auto-generated functions have three inputs,
\begin{equation}
\texttt{(tensors, which{\_}net, which{\_}env)} \nonumber
\end{equation}
and always return a single tensor, corresponding to the contraction of one of the networks in the project file it was generated from. The three input variables are described below.

\begin{figure}[!t!b]
\begin{center}
\includegraphics[width=6.5cm]{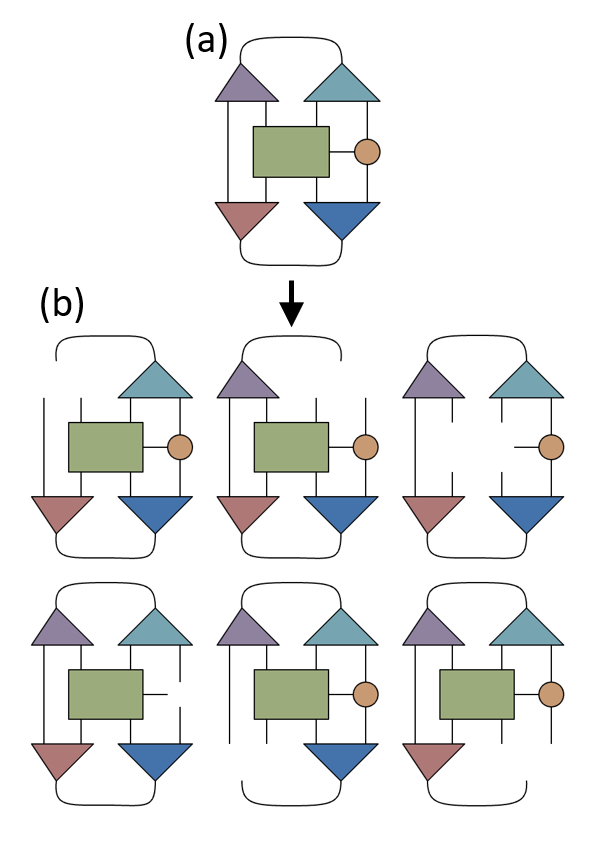}
\caption{(a) An example of a closed tensor comprised of six tensors. (b) Depiction of the six single tensor environments generated from the example closed network, whose contraction code can be automatically generated by \emph{TensorTrace}.} \label{fig:envs}
\end{center}
\end{figure}

\begin{itemize}
\item \texttt{tensors}: should be a cell array (MATLAB) or list array (Python and Julia) containing the unique tensors within the project file, ordered according to their assigned network numbers (this order is also given in the header of the function file). In the example ``binaryMERA" project of Fig. \ref{fig:binaryMERA3} there are four unique tensors, such that variable ``\texttt{tensors}'' is defined:
\begin{equation}
\textrm{\texttt{tensors = \{uDis, wIso, hamThree, rhoThree\}}} \nonumber
\end{equation}

\item \texttt{which{\_}net}: is an optional argument that specifies which network ($1-4$) within the project to evaluate (and defaults to the first valid network in the project if omitted from input).

\item \texttt{which{\_}env}: is an optional argument that can be used to specify the evaluation of single tensor environments from a closed tensor network (i.e. one that evaluates to a scalar). Setting \texttt{which{\_}env = 0} (or omitting the argument from input) will just evaluate the specified network as it appears. However, setting \texttt{which{\_}env} to a positive integer $M$ will instead contract for the single tensor environment (or network derivative) of the $M^\textrm{th}$ tensor of the specified network, see Fig. \ref{fig:envs} for an example. These environments are computing using the results from Ref. \onlinecite{NCON3}, which provides an efficient means to contract for any of the environments once the optimal contraction order for closed network is known.  
\end{itemize}
The complete example of how the auto-generated code can be employed in the full algorithm\cite{MERA2,MERA3} to optimize a binary MERA is provided with the \emph{TensorTrace} download (in the ``script{\_}binaryMERA" files). Although the auto-generated function files are most easily used directly, as described above, they could also be dissected to obtain the lines of code pertaining to individual network contractions. This follows as, internally, the function files simply contain a set of calls to the ``ncon" network contractor\cite{NCON1}, each of which is a few lines of code, which may be individually extracted. The role of \emph{TensorTrace} can thus be understood as that of an interpreter: it translates the network diagrams, and the information about their optimal contraction order, into a standard format (here the format used by ``ncon") which is then passed to another program which performs the actual network contractions. Thus it would also be relatively easy to employ the \emph{TensorTrace} auto-generated code with a different tensor contraction toolbox\cite{TB1,TB2,TB3,TB4,TB5,TB6} other that ``ncon", as this would simply require the network information to be translated to the appropriate conventions of another toolbox.

\section{Outlook}
We envision that there are several areas in which \emph{TensorTrace} could be useful to practitioners (or would-be practitioners) of tensor network methods:

\begin{itemize}
\item \textbf{Algorithm development:} in the design of new tensor algorithms \emph{TensorTrace} could provide quick evaluation of the computational costs associated to contracting a network, allowing the user to easily determine of feasibility of contraction and the maximum bond dimensions that would be viable. The software could also be used for rapid prototyping and reliable testing of new ideas.

\item \textbf{Algorithm implementation:} we envision that \emph{TensorTrace} would be useful in facilitating the implementation of existing tensor algorithms by substantially reducing the labor involved and removing potential sources of programming bugs. This could be particularly useful for novices to the field of tensor networks where, even with a proper and complete understanding of a given tensor algorithm, numerical implementations remain a formidable task due to the complexity of correctly programming tensor contractions. 

\item \textbf{Code maintenance:} storing a \emph{TensorTrace} project file alongside a numerical code would allow the code to be easily understood, maintained and modified at a later date (and by a third party, if desired). Otherwise, interpreting (or reinterpreting) the functioning a tensor network algorithm from the code alone is usually exceptionally difficult.

\item \textbf{Manuscript presentation:} although not the main purpose of \emph{TensorTrace}, it can export tensor diagrams into PNG figures which can then be used in manuscripts. One benefit of this approach is that it ensures that the manuscript depiction of an algorithm exactly matches its numerical implementation (since both are generated from the same project file). All of the tensor diagrams presented within the current manuscript were exported from \emph{TensorTrace}.  
\end{itemize}
The \emph{TensorTrace} software is still in an early stage of development; comments, bugs reports and suggestions for improvement are welcome (via the \emph{TensorTrace} website). The author would like to thank Adil Gangat and Wangwei Lan for testing and feedback.

\end{document}